# On what Terms and why the Thermodynamic Properties of Polymer Solutions Depend on Chain Length up to the Melt.

A. SCHNEIDER,[1] N. SCHULD,[1] M. BERCEA,[2] B. A. WOLF[1]

[1] Institut für Physikalische Chemie der Johannes Gutenberg-Universität Mainz
Jakob-Welder-Weg 13, D-55099 Germany

[2] Permanent address: "Petru Poni" Institute of Macromolecular Chemistry,
Iaşi, Romania

**ABSTRACT:** Theoretical considerations based on chain connectivity and conformational variability of polymers have lead to an uncomplicated relation for the dependence of the Flory-Huggins interaction parameter $\chi$ on the volume fraction of the polymer, $\varphi$, and on its number of segments, $N$. The validity of this expression is being tested extensively by means of vapor pressure measurements and inverse gas chromatography (complemented by osmotic and light scattering data from literature) for solutions of poly(dimethylsiloxane) in the thermodynamically vastly different solvents n-octane (n-$C_8$), toluene (TL), and methylethylketone (MEK) over the entire range of composition for at least six different molecular masses of the polymer. The new approach is capable to model the measured $\chi(\varphi, N)$ very well, irrespective of the thermodynamic quality of the solvent, in contrast to traditional expressions, which are often restricted to good solvents but fail for bad ones and *vice versa*. At constant polymer concentration the $\chi$ values result lowest for n-$C_8$ (best solvent) and highest for MEK (theta solvent); the data for TL fall between. The influences of $N$ depend strongly on the thermodynamic quality of the solvent and are not restricted to dilute solutions. For good solvents $\chi$ increases with rising $N$. The effect is most pronounced for n-$C_8$, where the different curves for $\chi(\varphi)$ fan out considerably. The influences of $N$ become less distinct for TL, and for MEK they vanish at the (endothermal) theta temperature. For worse than theta conditions, the $\chi$ values of the long chains become less than that of short ones. This change in the sign of $N$-influences is in agreement with the present concept of conformational relaxation.

## INTRODUCTION

The determination of Flory-Huggins interaction parameters, $\chi$, for polymer/solvent systems via their equilibrium vapor pressures has become much easier during the past years because of an automated method that combines head space sampling and gas chromatography[1] By means of the additional experimental material acquired in this manner it has become obvious that the molecular weight dependence of $\chi$ is in contrast to general belief not restricted to dilute polymer solutions, but expands up to vol-

Correspondence to: B.A. Wolf (E-mail: Bernhard.Wolf@Uni-Mainz.de)

ume fractions $\varphi$ of the polymer exceeding 0.8 under certain conditions. In the preceding contribution[2] we have studied how the solvent quality controls the concentration and temperature dependence of $\chi$. The present investigation was performed to investigate the molecular weight influences on $\chi$ in the region of high chain overlap. Poly(dimethylsiloxane) was again chosen as the polymer component of the different binary systems because of its low glass transition temperature. The interpretation of the present results is performed by means of a new concept[3,4], which is based on the connectivity of the segments of a polymers chain and on the conformational variability of macromolecules of the present type. Among other things this approach predicts a change in the sign of the influences of the molar mass M of the polymer on $\chi$, as the solvent quality falls below that of theta solvents. This is the reason why we have extended our measurements into this area.

## EXPERIMENTAL PART

### Substances

Six of the poly(dimethylsiloxane) samples with molecular weights (number average) between 6 kg/mol and 150 kg/mol were commercial products of Wacker, Germany. They were dried at 80 °C in a vacuum oven for 2 days before use. The higher molecular weight samples were synthesized as described in literature.[3] Molecular weights stem from GPC experiments. They are listed in Tab. 1 together with other relevant data. The solvents were of p.a. grade and used without further purification. The gases needed for gas chromatography (GC) were helium 4.6 (Messer-Griesheim, Germany) and dried air. Nitrogen was used as carrier and methane as non-interacting marker in case of the IGC-measurements, where Chromosorb W (AW-DMCS treated, 80/100 mesh, SIGMA-Aldrich) was utilized as support.

Tab. 1: Characterization of the poly(dimethylsiloxane) (PDMS) samples and solvents used

| Acronym | $M_n$ [kg /mol] | U | Supplier | Density $\dfrac{\rho}{g\,cm^{-3}}$ with $\vartheta$ / °C |
|---|---|---|---|---|
| PDMS | 6 | 0.82 | Wacker | $\dfrac{0.97}{1+9.2\cdot 10^{-4}(\vartheta-25)+4.5\cdot 10^{-7}(\vartheta-25)^2}$ |
|  | 52 | 0.58 |  |  |
|  | 73 | 0.67 |  |  |
|  | 87 | 0.75 |  |  |
|  | 100 | 0.80 |  |  |
|  | 150 | 0.86 |  |  |
|  | 279 | 0.8 | Synthesized as described in ref[3] |  |
|  | 317 | 0.7 |  |  |
|  | 444 | 0.8 |  |  |
|  | 525 | 0.58 |  |  |
|  | 605 | 0.8 |  |  |
|  | 865 | 3 |  |  |
| MEK |  |  | Fluka | $0.82592 - 0.00106\cdot\vartheta$ |
| TL |  |  |  | $0.8854 - 0.00093\cdot\vartheta$ |
| n-$C_8$ |  |  |  | $7185 - 8.239\cdot 10^{-4}\vartheta + 4.459\cdot 10^{-7}\vartheta^2 - 5.293\cdot 10^{-9}\vartheta^3$ |

**Headspace gas-chromatography (HSGC)**

This method is extensively described in literature.[2] It uses a headspace sampler in combination with a gas chromatograph. The removal of air from the polymer solution is not required since its components are separated from the peaks of interest by the GC (TC-detector). The concentrations of the samples were chosen in a region between 40 and 95 wt% polymer. Because of the different available amounts of the polymers (only 7-20 g of the high molecular weight polymers) two different methods of determining the vapor pressure were applied: Normally one extracts four to six times a defined amount of the vapor above the polymer solution, leaving enough time between each removal (approx. 12 to 24 hours) to reestablish the equilibrium. This method requires a sufficiently large reservoir of solvent so that the different extractions do not change the polymer concentration significantly. The second method, was developed by analogy to a procedure called "multiple headspace extraction", (MHE[5]) This variant is particularly suited for small samples and such with low content of solvent. Again a defined amount of vapor is taken, up to six times, but with only a short time interval (10 to 15 min) between each extraction. In this case the amount of solvent decreases from extraction to extraction. In order to obtain the equilibrium vapor pressure for the initial composition, the logarithm of the measured peak area is plotted as a function of the number of extractions and the value extrapolated for the first extraction is taken instead of the measured value, which is according to experience insecure. This method was used for the higher molecular weight samples because only approximately 0.5 mL solution is required per concentration.

**Inverse gas chromatography (IGC)**

All columns for IGC-measurements were prepared in the same way: A solution of the PDMS sample of interest in chloroform was prepared and a calculated amount of Chromosorb W was added. The main part of the solvent was removed by evaporation under stirring. After that the solid support was dried for 2-3 days by 50 °C in vacuo. A stainless steel column (100 cm long) was packed with the so prepared material by applying vacuum to one end. Finally the column was closed by glass wool on both ends. The details of the loading and the columns used for the IGC measurements are collected in Tab. 2.

Tab. 2: Specification of the IGC columns used for the determination of $\chi_\infty$

| Polymer | Loading (wt.%) | Wt polymer (g) | Column length (cm) |
|---|---|---|---|
| PDMS 279n | 9.2 | 0.4849 | 100 |
| PDMS 444n | 9.1 | 0.4871 | 100 |
| PDMS 865n | 9.2 | 0.4988 | 100 |

The inverse gas chromatography measurements were performed by means of a Sygma 300 Perkin-Elmer gas chromatograph equipped with a flame ionization detector. The column temperature was controlled to ± 0.2 °C over the measured temperature range (25-110 °C). For the lower temperatures (25-60 °C) an additional water bath was used to keep the temperature constant. The columns were kept at 80 °C over night before use, flushing them with the carrier gas nitrogen. Methane, as a non interacting marker, was used to correct for the dead volume of the column. The retention time was determined by an Olivetti M-24 microcomputer. At least 3 measurements were carried out for each polymer sample and temperature. Inlet and outlet pressures of the column were measured by a mercury manometer and the flow rates, determined at the end of the column with a bubble flowmeter, were adjusted between 5 and 9 mL/min. The solvents (< 0.01 µL, including a small amount of the marker) were injected manually with a 10 µL Hamilton syringe. More details can be found in the literature.[6]

**THEORETICAL BACKGROUND**

During the last decade it has become clear beyond doubt[1] that the Flory-Huggins interaction parameter $\chi$ may still vary considerably with chain length even at the highest polymer concentrations. These observations contradict one of the central theoretical postulates, namely the loss of "individuality" of the solute molecules beyond the so called chain overlap concentrations. In order to rationalize this additional experimental information and to incorporate it into existing theories, we have treated the new phenomena in terms of the chain connectivity and conformational variability of polymers.[3,4] These considerations have lead to the following straightforward expression for the variation of $\chi$ with the volume fraction, $\varphi$, of the polymer.

$$\chi = \frac{\alpha}{(1-\nu\varphi)^2} - \zeta(\lambda + 2(1-\lambda)\varphi) \quad (1)$$

The parameter $\alpha$ measures the contribution of the insertion of solvent molecule between two contacting polymer segments in the range of pair interaction between the solute, without changing the conformation of the two polymer chains engaged in this process. In case the Gibbs energy of the total system can be lowered by a rearrangement of these chains in the vicinity of the new contacts their conformation will change. The contribution of this transaction is quantified by the parameter $\zeta$. In the limit of $\varphi \to 0$ the above relation reduces to

$$\chi_o = \alpha - \zeta\lambda \quad (2)$$

where $\chi_o$ is related to the second osmotic virial coefficient $A_2$ as[7]

$$\chi_o = \frac{1}{2} - A_2 \rho_2^2 \overline{V_1} \quad (3)$$

$\rho_2$ being the density of the polymer and $\overline{V_1}$ the molar volume of the solvent.

The following relation[3] holds true for $\lambda$

$$\lambda = \left(\frac{1}{2} + \kappa N^{-(1-a)}\right) \quad (4)$$

where $\kappa$ is given by

$$\kappa = K\rho_2 \left(\frac{\rho_2}{\rho_1} M_1\right)^a \quad (5)$$

and $N$ represents the number of the polymer segments defined in the usual manner. $K$ and $a$ stand for the parameters of the Kuhn-Mark-Houwink relation; $M_1$ and $\rho_1$ are the molar mass and the density of the solvent. Finally, $\nu$ accounts for the facts that the deviation from combinatorial behavior becomes smaller as the degree of coil overlap increases and that the surfaces of the components are more adequate for the description of the interactions between them than their volumes.

Four parameters are according to eq (1) required to describe $\chi(\varphi)$. In view of the fact that $\lambda$ is only slightly larger than 0.5 for high values of $N$, we can merge $\zeta$ and $\lambda$ into only one parameter by rewriting this relation as

$$\chi = \frac{\alpha}{(1-\nu\varphi)^2} - \zeta\lambda\left(1 + 2\left(\frac{1}{\lambda}-1\right)\varphi\right) \qquad (6)$$

and setting

$$\frac{1}{\lambda} - 1 \approx 1 \qquad (7)$$

Furthermore, in view of the fact that $\chi_0$ is easily accessible (e.g. from light scattering or osmotic experiments), it is advantageous to rewrite (6) by means eq (2) as

$$\chi = \frac{\chi_0 + \zeta\lambda}{(1-\nu\varphi)^2} - \zeta\lambda(1+2\varphi) \qquad (8)$$

The thus obtained relation contains only two adjustable parameters ($\zeta\lambda$ and $\nu$), if $\chi_0$ is known with sufficient accuracy.

Eq (8) enables a very accurate description of the huge diversity of the experimentally obtained dependencies $\chi(\varphi)$.[4] Fig. 1 shows its application to the data of the preceding publication[2], which were so far evaluated in the conventional manner, i.e. either in terms of a series expansion or – if possible – according to the relation of Koningsveld and Kleintjens.[8] The adjusted parameters are collected in Tab. 3.

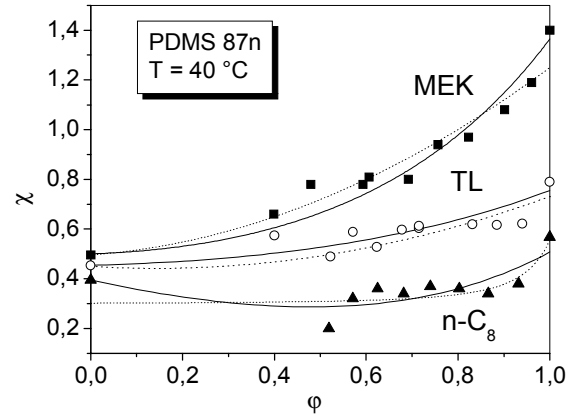

**Fig. 1:** Flory-Huggins interaction parameters $\chi$ as a function of the volume fraction $\varphi$ of poly(dimethylsiloxane) ($M_n$ = 87 kg/mol) in different solvents at 40 °C. MEK: methylethylketone; TL: toluene: n-C$_8$: n-octane. Full lines: description by means of eq (8); dotted lines: previous representation.[2]

Table 3: Parameters of eq (8) describing the experimental data shown in Fig. 1

|      | $\alpha$ | $\zeta \cdot \lambda$ | $\nu$ |
|------|----------|-----------------------|-------|
| MEK  | 0.632    | 0.136                 | 0.385 |
| TL   | 0.609    | 0.155                 | 0.294 |
| n-C8 | 0.895    | 0.500                 | 0.332 |

In the following section we report extensive data on $\chi(\varphi)$ obtained for PDMS samples of very different molar masses in three solvents of maximum possible variation in thermodynamic quality. The results are presented in the order of declining solvent power.

**RESULTS AND DISCUSSION**

The first evaluation of the experimental findings concerns the composition dependence of the Flory-Huggins interaction parameter for the different solvents and molar masses of the polymer. In order to ease a direct comparison of the results, the scale of the $\chi$-axes is in all graphs the same, namely that required for the worst solvent. After this analysis we discuss the influences of chain length on $\chi$ for three different composition ranges: Infinitely dilute solutions ($\chi_o$), 80 vol% polymer content ($\chi_{0.8}$) and infinitely concentrated ($\chi_\infty$).

**Concentration dependence**

*n-Octane*

This solvent is to our knowledge the very best for PDMS. A selection of experimental data obtained from HSGC, light scattering ($\chi_o$) and IGC ($\chi_\infty$) for nine polymer samples of different molar mass is shown in Fig. 2, together with the modeling of these results according to eq (8).

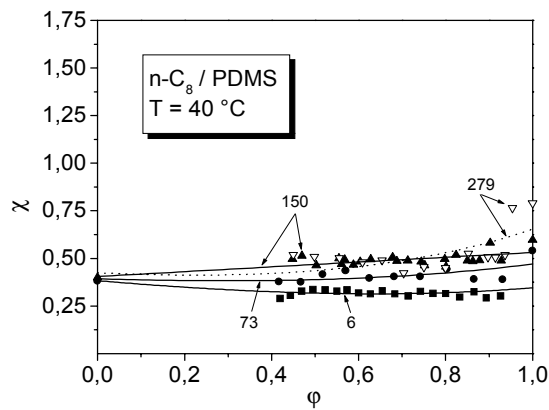

**Fig. 2:** Composition dependence of $\chi$ for solutions of PDMS in n-C$_8$ at 40 °C. The number average molar masses (in kg/mol) are indicated in the graphs. The lines are calculated by means of eq (8). The data points for 150 and 297 kg/mol are very similar except for polymer concentrations larger than 0.85 wt%. The adjustment for the latter polymer is shown as a dotted line.

In agreement with earlier observations[2], the concentration dependence of $\chi$ is only moderate for this extremely good solvent in comparison with worse conditions, as for instance realized with theta solvents. Another typical feature of very good solvents consists in negative initial slopes of $\chi(\varphi)$ in combination with a possible increase of $\chi$ in the limit of the pure melt. This situation leads to more or less pronounced minima in this dependence. The curves for the different molar masses of the polymer fan out significantly in the case of good solvents, as expected from the general concept of conformational response. Under these conditions this effect persists up to the pure melt. Beyond experimental uncertainty IGC yields $\chi$ values which increase with rising chain length.

## Toluene

Although somewhat worse in thermodynamic quality than n-$C_8$, TL is still a very good solvent as can be seen (Fig. 3) from the fact that the $\chi$ values remain comparatively low, irrespective of composition. This time there is no indication for the occurrence of minima in $\chi(\varphi)$. According to the somewhat lower quality of TL the curves for the different molar masses do not fan out as much as for n-$C_8$. Furthermore, the IGC data result less dissimilar than for TL, as expected according to the present approach.

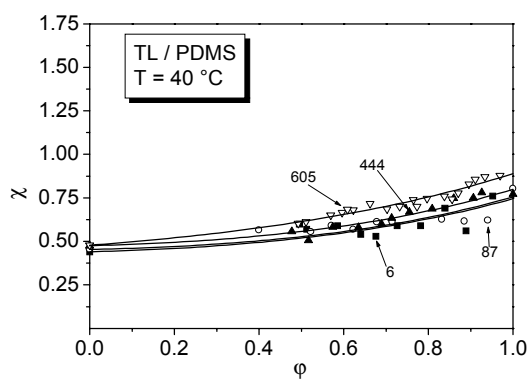

**Fig. 3:** As Fig. 2, but for TL as the solvent component. Again the $\chi$ values increase markedly as the molar mass of the polymer rises.

## Methylethylketone

MEK exhibits the least solvent power for PDMS. It becomes a theta solvent[9] at 20 °C and phase separation may set in as the temperature falls below that value. Fig. 4 depicts the results for this system at three temperatures.

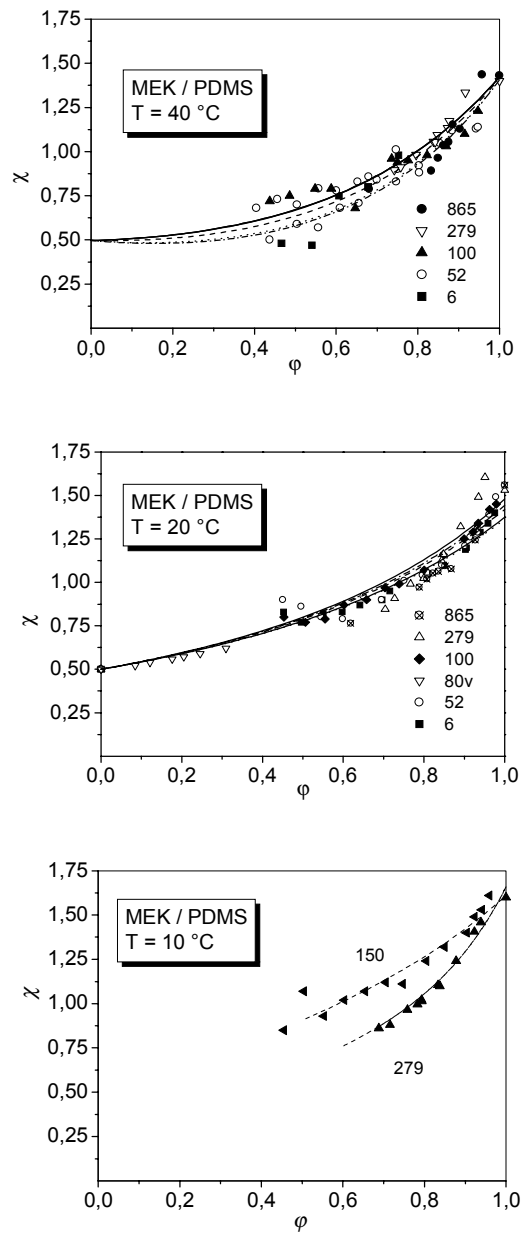

**Fig. 4:** As Fig. 2, but for the solvent MEK (theta temperature 20 °C).
*Upper part:* 40 °C; for this evaluation only two the parameters ($\zeta\lambda$ and $\nu$) were adjusted (known $\chi_0$ values); for calculation of the curves we have used the average value of $\nu$ (=0.415) because of its little variation with chain length.
*Middle part:* 20 °C; due the fact that this is the $\Theta$ temperature, $\zeta\lambda = 0$ and $\chi_0 = 0.5$, irrespective of molecular weight. The data 80v were taken from the literature.[10]
*Lower part:* 10 °C; for worse than theta conditions the $\chi$ values for the shorter chains become larger than that for the longer chains.

At 40 °C (Fig. 4a) the influence of molar mass on $\chi$ is qualitatively still the same as for good solvents, i.e. the interaction parameter rises upon an augmentation of $M$. For 20 °C (Fig 4b), the theta temperature of the system, $\chi$ is within experimental error not influenced by the chain length of the polymer over the entire range of composition. This observation is consistent with the present approach, according to which conformational relaxation, the reason for $\chi(N)$, should have no effect. At $T = \Theta$ the polymer chains remain unperturbed from the infinitely dilute state up to the pure melt. In the case of 10 °C phase separation will already take place for the higher molecular weight polymer samples in mixtures with high solvent content. The experimental information (Fig. 4c) required for the discussion of $\chi(M)$ is therefore limited to high polymer concentrations at which the system is still homogenous, irrespective of $M$. The obtained data (for the sake of clarity only shown for two molecular weights) demonstrate that the influence of chain length exhibits the opposite sign under these very unfavorable thermodynamic conditions. Now the interaction parameter *decreases* with rising $M$. This finding is predicted by the present approach. The conformational response $\zeta$ changes its sign as the solvent power falls below theta conditions.[3,4] Under these circumstances the formation of intersegmental contacts is so strongly favored that the coils shrink upon dilution.

**Chain length dependence**

The influence of the molar mass of the polymer on the Flory-Huggins interaction parameters will be discussed for three different composition ranges. In addition to the limiting situations ($\chi_o$, vanishing polymer content, and $\chi_\infty$, vanishing solvent content) we pay particular attention to highly concentrated polymer solutions and test, how $\chi$ varies with $M$ in the case of a constant volume fraction of the polymer equal to 0.8 ($\chi_{0.8}$) in order to control the validity of the present approach. An evaluation of the data in terms of a power law was disregarded, because of the unrealistic limiting behavior for infinitely long chains, despite the fact that $\ln \chi$ depends in very good approximation linearly on $\ln M$ for all systems and concentrations under investigation. In place of the traditional analysis the discussion is performed on the basis of eqs (6) and (4), leading to finite interaction parameters for $M \rightarrow \infty$.

*$\chi_o$: Vanishing polymer content*

The accuracy of the experimental information obtained for this composition range is uncommonly high due to the fact that the directly accessible second osmotic virial coefficients measure the difference $0.5 - \chi_o$ (eq (3)); this fact is acting as a "magnifying glass" for the effects of dilution. In this concentration range the particularities of a given polymer/solvent system become most obvious. The reason is that the deviations of the actual coil dimensions from the unperturbed dimension, prevailing in the pure polymer melt, reach their maximum. The local effect of conformational relaxation is under these conditions affecting the entire polymer molecules. Consistent with eqs (2) and (4) of the present approach, data evaluation is performed by plotting $\chi_o$ as a function of $M^{-(1-a)}$ as shown in Fig. 5.

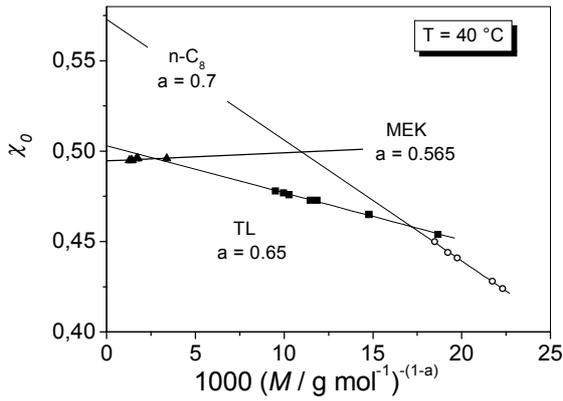

**Fig. 5:** Dependence of $\chi_o$ on the molar mass $M$ of the polymer evaluated according to the eqs (2) and (4) for different solvents and 40 °C; $a$, the exponent of the Kuhn-Mark-Houwink relation, characteristic for the different polymer/solvent systems is indicated in the graph.

It is normal that the dependencies resulting for different solvents or temperatures intersect each other[11] for reasons discussed below. The slope of the lines shown in Fig. 5 is according to eqs (2) and (4) proportional to $-\zeta$. Due to the fact that $\zeta$ constitutes a quantitative measure of solvent quality, the steepness increases from MEK via TL to n-$C_8$. This finding is in agreement with intuition. The observed intercepts, on the other hand, surprise at first, because one would expect that the limiting value of $\chi_o$ for infinitely long chains should reflect the solvent quality in the usual way. In other words one anticipates the least value for n-$C_8$ and the largest for MEK, but just the opposite is the case.

The following considerations make the finding comprehensible. For the limiting value of $\chi_o$ in the case of infinite molar mass of the polymer the present approach yields the expression

$$\lim_{N \to \infty} \chi_o = \alpha - \frac{\zeta}{2} \qquad (9)$$

It states that the value of the intercept is determined by the coaction of two parameters, which vary inversely with solvent quality. A clear-cut dependence on solvent quality can, however, be established if one bears in mind that the parameters $\alpha$ and $\zeta$ cannot be altered independently. According to experimental evidence[3] there exists in very good approximation a linear correlation, which reads

$$\alpha = B\zeta + \frac{1}{2} \qquad (10)$$

where $B$ is constant for a given system. Inserting this relation into eq (2) yields

$$\chi_0 = \zeta(B - \lambda) + \frac{1}{2} \qquad (11)$$

from which we obtain the following expression for the limiting case of infinitely long chains by means of eq (4)

$$\lim_{N \to \infty} \chi_o = \zeta\left(B - \frac{1}{2}\right) + \frac{1}{2} \qquad (12)$$

Because of the fact[3] that $B$ is larger than 0.5 in all cases, the expression in the brackets of eq (12) remains positive. This implies that the intercept of the different lines of Fig. 5 with the ordinate, the Flory-Huggins interaction parameter for infinitely long chains, increases with rising solvent quality. This conclusion is well validated by the present findings and explains, why the lines for different solvents intersect.

### $\chi_{0.8}$ : Highly concentrated solutions

For such highly concentrated polymer solutions the chain overlap has beyond doubt become large enough to result in unperturbed dimensions. Under these conditions chain expansion or shrinking can no longer result in a lowering of the Gibbs energy of the system because the reservoir of pure solvent has already been emptied long before that concentration is reached. Consequently the exponent of the generalized Kuhn-Mark-Houwink equation, relating the specific volume of polymer coil in a mixture of given composition to the molar mass of the polymer, becomes 0.5, i.e. assumes the value for unperturbed (theta) conditions. The evaluation of experimental data is therefore in all cases performed with $a = 0.5$, irrespective of solvent quality (Fig. 6).

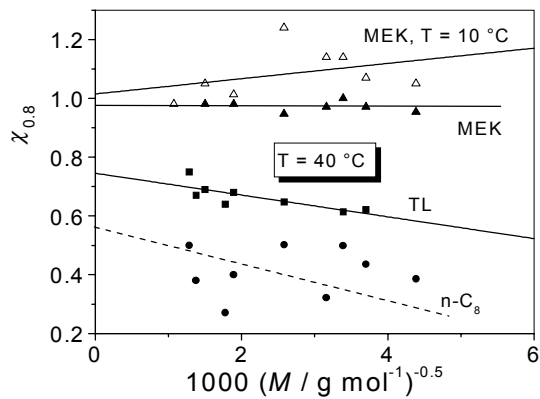

**Fig. 6:** Dependence of the Flory-Huggins interaction parameter on the molar mass $M$ of the polymer for $\varphi = 0.8$ at the indicated temperatures. Due to the large extent of coil overlap under these conditions $a$ is set equal to 0.5 for all solvents. The broken line is in contrast to all others a guide for the eye only.

The generally neglected persistence of the influences of chain length up to extremely large polymer concentrations has already been reported.[1] The present findings allow some more detailed insight, in particular concerning the influence of the thermodynamic quality of the solvent on these effects. In agreement with the conclusions based on the present approach, $\chi_{0.8}$ may vary considerably with $M$. For TL, a good solvent, the interaction parameters rise markedly as the chains become longer. With the extremely good solvent n-$C_8$ the scattering of the experimental data is for all experiments much higher than usual. The reason for that behavior is as yet unclear, it might have to do with the high flexibility of the polymer chain and the possibility to form helical structures in certain solutions. Despite these uncertainties, the trend of $\chi(M)$ is clear and the data are compatible with the theoretical postulate that the effects should be the more pronounced the better the solvent becomes. Further evidence for the validity of the present approach is given by the change in the sign of the dependencies shown in Fig. 6. As the solvent becomes very poor, i.e. as its quality falls below that of a theta solvent, the conformational response to dilution does no longer lead to more spacious conformations but to more closely packed arrangements because of the large preference of intersegmental contacts over contacts between solvent molecules and polymer segments.

### $\chi_\infty$: *Vanishing solvent content*

The difficulty in obtaining exact data for the interaction between a probe molecule of the solvent and the segments of a polymer melt from IGC measurements are comparable or even larger than that discussed in the last section for highly concentrated polymer solutions. Despite this fact some trends are clearly noticeable as can be seen from Fig. 7, where $a$ was again set 0.5 (cf. last paragraph).

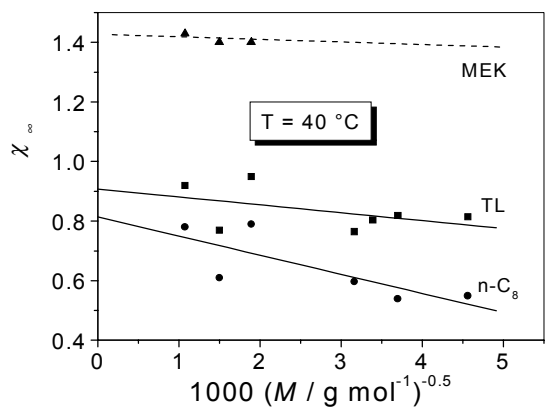

**Fig. 7**: Dependence of the Flory-Huggins interaction parameter on the molar mass $M$ of the polymer for $\varphi \to 1.0$ at 40 °C. For TL and n-C$_8$ the data shown for the three samples of lowest molar were taken from the literature.[12]

According to the present ways of thinking there is no reason why the effects observed for highly dilute or moderately concentrated polymer solutions should fade away in the limit of vanishing solvent concentration. In view of the fact that the persistence of the chain length dependence of $\chi$ up to the melt may seem counter intuitive, some additional comments appear advisable. First of all, it should be irrelevant that the fraction of solvent approaches zero under these conditions, because $\chi$ is normalized to one mole of solvent. Still it is hard to imagine why the chain length of the polymer should influence $\chi$ even in the absence of special effects resulting from end-groups. Within the scope of the present approach this behavior results from the contribution of the conformational relaxation to the Flory-Huggins interaction parameter. That part is proportional to $\chi_{\text{isol. coil}}$ (quantifying the opening of an intramolecular contact between two segments) and varies with chain length.[3] The easiest way to rationalize the findings is offered by a somewhat different but equivalent description of the situation in terms of individual parameters for inter- and intramolecular interactions, which are in the general case not equal.[11] The present observation can then be explained by a chain length dependence of the mixing ratio of these two parameters.[11] Indeed the probability to open an intermolecular contact results larger than 0.5 for low $N$ values (in case of oligomers this value is close to unity) and decreases steadily as $N$ rises until it reaches 0.5 in the limit of $N \to \infty$.

**CONCLUSIONS**

The composition and chain length dependence of the Flory-Huggins interaction parameter resulting from a recent approach based on chain connectivity and conformational variability is confirmed by the present experiments. In contrast to most customary relations, which are not generally valid, it can model $\chi(\varphi)$ for any thermodynamic quality of the solvent. Furthermore the approach predicts three features, which have all been confirmed experimentally. (i) The extent to which $\chi$ depends on $N$ is governed by the particular thermodynamic situation. The best solvents exhibit the largest increase of $\chi$ with rising $N$. (ii) The effects of chain length are preserved over the entire composition range. (iii). As the solvent quality falls below theta conditions, the influence of $N$ on $\chi$ changes its sign; in this case the interaction parameters decrease as the molecular weight rises.

**Acknowledgements:** Our special thanks go to the DFG for continued financial support. Furthermore we are particularly grateful to Prof. Iruin and his coworkers for their help with the IGC investigation.